\documentclass[showkeys,superscriptaddress,notitlepage]{revtex4-1}
\usepackage{graphics,graphicx} 
\usepackage{geometry}
\usepackage{epstopdf}
\usepackage{natbib}
\begin{document}

\title
{Absence makes the heart grow fonder: social compensation when failure to interact risks weakening a relationship}

\author{Kunal Bhattacharya}
\email[Corresponding author; ]{kunal.bhattacharya@aalto.fi}
\author{Asim Ghosh}
\author{Daniel Monsivais}
\affiliation {Department of Computer Science, Aalto University School of Science, P.O. Box 15400, FI-00076 AALTO, Finland}
\author{Robin Dunbar}
\affiliation {Department of Experimental Psychology, University of Oxford, South Parks Rd, Oxford, OX1 3UD, United Kingdom}
\affiliation {Department of Computer Science, Aalto University School of Science, P.O. Box 15400, FI-00076 AALTO, Finland}
\author{Kimmo Kaski}
\affiliation {Department of Computer Science, Aalto University School of Science, P.O. Box 15400, FI-00076 AALTO, Finland}
\affiliation {Department of Experimental Psychology, University of Oxford, South Parks Rd, Oxford, OX1 3UD, United Kingdom}

\begin{abstract}
Social networks require active relationship maintenance if they are to be kept at a constant level of emotional closeness. For primates, including humans, failure to interact leads inexorably to a decline in relationship quality, and a consequent loss of the benefits that derive from individual relationships. As a result, many social species compensate for weakened relationships by investing more heavily in them. Here we study how humans behave in similar situations, using data from mobile call detail records from a European country. For the less frequent contacts between pairs of communicating individuals we observe a logarithmic dependence of the duration of the succeeding call on the time gap with the previous call. We find that such behaviour is likely when the individuals in these dyadic pairs have the same gender and are in the same age bracket as well as being geographically distant. Our results indicate that these pairs deliberately invest more time in communication so as to reinforce their social bonding and prevent their relationships decaying when these are threatened by lack of interaction.
\end{abstract}

\maketitle

\section*{Introduction}

For intensely social species like primates, close friendships buffer the individual against the stresses of living in enforced proximity to others, and so make living in social groups possible \cite{van1983diurnal,Dunbar1837}.  In both primates \cite{silk2003social,cameron2009social,silk2010strong} and humans \cite{uchino2006social,spence1954one,smith2008social,oesch2015influence}, the number and quality of a female's social relationships has a direct effect on the stress she suffers, the illness she and her offspring experience, the number of offspring she has and even her survival (see also \cite{holt2010social}). However, in both anthropoid primates and humans, such relationships depend on frequent interaction, often at quite specific rates, to maintain relationship quality \cite{lehmann2007group,roberts2015managing}. In humans at least, interacting at less than the specified frequency results in the rapid decline in the emotional quality of a relationship \cite{saramaki2014persistence}, and hence the relationship's effectiveness as a buffer. Primates, and humans, thus invest considerable quantities of time in maintaining relationship quality. 
Perhaps because relationships are so important, animals seem to be sensitive to fact that relationship quality may be adversely affected by circumstances. Baboon mothers, for example, are obliged by the foraging demands of lactation to reduce time invested in grooming their main social partners; they later exhibit a rebound effect in which they groom these individuals more than usual once the infant has started to wean and requires less of her time to be devoted to feeding \cite{altmann2001baboon,dunbar1988maternal}. Similarly, greetings are often more intense when animals have been separated for some considerable time. Elephant greetings, for example, are more elaborate following a prolonged absence \cite{moss1983relationships}, while among bonobos joiners who had been separate from residents longer receive the most socio-sexual solicitations \cite{wittig2005repair}. Similar findings have been noted following conflicts.: in both primates \cite{moscovice2015welcome,aureli2002conflict} and hyenas \cite{smith2011greetings}, greeting behaviour is more intense when the past interactions between individuals have been aggressive. In sum, it seems that it takes more effort to repair damaged ties than to maintain existing stable ties.
	Humans are no less susceptible to the risks imposed by the fragility of  close relationships, and may be expected to adopt similar strategies to manage their relationships. However, the inelasticity of time invariably restricts the effort an individual can invest such that only a certain number of relationships would be stable at different levels of emotional intensity \cite{hill2003social,sutcliffe2012relationships,zhou2005discrete}. Hence, while humans obviously maintain large and complex social networks \cite{hill2003social,dunbar2016online}, constraints of cognition and time play a vital role in shaping how well their social interactions function \cite{nie2001sociability,roberts20106}. If investment effort is insufficient, social ties will inexorably decay \cite{burt2000decay,burt2002bridge}. Even though the advent of new technologies has introduced a number of new communication channels, the maintenance of relationships still requires investment of considerable time and effort \cite{oswald2003best,cummings2006communication,miritello2013time}. Nonetheless, these new technologies have one key advantage, namely allowing individuals to interact and service a relationship when they are geographically far apart.

In this paper, we use a large dataset from a mobile phone service provider in a European country, spanning over seven consecutive months in the year $2007$, to investigate whether humans adjust their investment effort to compensate for reduced contact. To do this, we first examine whether the duration of calls is related to the time since last contact, and then ask whether this is especially true of individuals who are geographically further apart (and hence cannot meet up in person so easily). For a given ego (individual) we study the data on  mobile phone communication with the alters (individuals participating in calls with the ego) with whom the communication events are sufficiently spaced in time yet showing a degree of regularity. In the offline world, individuals who are contacted at least once a month fall into the category of close, as opposed to intimate, friends, and constitute the second most important group of network alters \cite{roberts20106,sutcliffe2012relationships}. As such they represent a stable group of alters that are emotionally important to ego, but ones that are not contacted so frequently as to obliterate any trace of the effect we are interested in. We expect that the investment by the ego in actively maintaining such ties would be reflected as a relationship between the inter-event times and the call durations. 

\begin{figure}[t]
\includegraphics[scale=1]{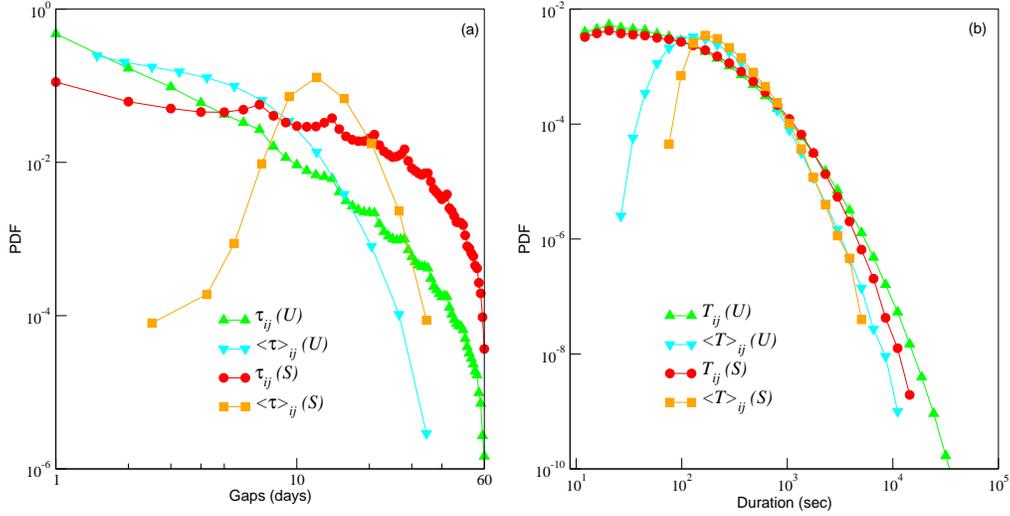}
\caption{(a) Probability distribution functions (PDFs) for inter-call times (gaps), $\tau_{ij}$ and PDFs for the average gaps, ($\langle \tau\rangle_{ij}$ -- calculated with respect to individual pairs $ij$). The pairs ($ij$s) are chosen irrespective of the age and gender of the individuals in the pair. The green up-triangles and blue down-triangles correspond to the set $\mathcal{U}$. The red circles and the orange squares correspond to the set $\mathcal{S}$. (b) PDFs for call durations, $T_{ij}$ and for the average call durations, ($\langle T\rangle_{ij}$). The different symbols are used in the same sense as in (a).}
\label{fig-1}
\end{figure}

\section*{Results}
\label{Results}

For a given ego-alter pair we measure the inter-event time (`gap') between two successive calls in number of days irrespective of the directionality of the calls. Then we examine the variation in the duration of the succeeding call (in seconds) as a function of the length of the inter-event time. For analytical convenience, instead of considering individual calls we aggregate all calls in a given day and ignore the days in which the total calling time is less than $10$ seconds. Thus by mentioning `call' or an `event' with respect to a given pair on a particular day  we refer to the aggregated voice communication on that day. For a given ego $i$ and its alter $j$, we construct the set of ordered pairs $\{(\tau_{ij},T_{ij})\}$, such that, $\tau_{ij}$ is a gap between two calls and $T_{ij}$ is the duration of the succeeding call. For a given pair $ij$, we define $\langle\tau\rangle_{ij}$ and $\langle T\rangle_{ij}$ as the average gap and the average duration of calls, respectively.

We concentrate on the ego-alter pairs for which the communication is sufficiently spaced over time by considering the set of pairs, $\mathcal{S}$, for which the maximum number of calls in any calendar month does not exceed $4$ and there is at least one call in each of the $7$ months. The bound on the maximum number of calls results in a characteristic gap of just over a week. Higher values of this bound would allow for more calls per week and would result in the inclusion of the frequently contacted alters for a given ego \cite{palchykov2012sex} and for such alters the probability of finding large gaps in communication would be comparatively less (illustrated in Fig. \ref{fig-1}). At least one call per month allows us to focus on relationships which may be considered otherwise stable. Additionally, we consider only those pairs for which the distance between their most common location ($d_{ij}$) is greater than zero to reduce the likelihood of face-to-face interaction. The detailed criteria for selecting pairs is provided in Materials and Methods and the robustness of the results is discussed in the Supplementary Material (SM). 

First we show the probability distribution of the gaps and the call durations corresponding to pairs belonging to the set $\mathcal{S}$. For comparison, we construct another set $\mathcal{U}$ by relaxing the restriction on the maximum number of calls per month (other parameters being the same as for $\mathcal{S}$) to include the more frequently contacted alters, such that $\mathcal{U}\supseteq \mathcal{S}$. In Fig. \ref{fig-1} we plot the probability distribution functions (PDFs) of $\tau_{ij}$ and $T_{ij}$ for the pairs in $\mathcal{S}$ and $\mathcal{U}$, irrespective of the age and gender of the individuals. In general, the PDFs for $\tau_{ij}$ and $T_{ij}$ are fat tailed. The PDF for $\tau_{ij}$ shows peaks at multiples of seven days, which indicates a high propensity to make calls during weekends. In general, the PDFs for the averages for individual pairs ($\langle\tau\rangle_{ij}$ and $\langle T\rangle_{ij}$)  show well defined peaks. Fig. \ref{fig-1} (a) and (b) shows that a typical average separation and a typical average call duration for alters in $\mathcal{S}$ is around $12$ days and $170$ seconds, respectively. The PDF  for the average gap in $\mathcal{U}$ falls off exponentially and the typical average duration in $\mathcal{U}$ is around $130$ seconds.
     
In Fig. \ref{fig-2} we plot the binned curves for the duration of the succeeding calls as a function of the gaps corresponding to communication between egos and the alters in the set $\mathcal{S}$. Each curve corresponds to an age and sex cohort for the ego and the sex cohort for the alter. The curves indicate a logarithmic increase in the duration of calls with the increase in the gap. Although, the behaviour is found across all the cohorts considered, within the ego-age range of $25-60$ year olds the trend appears to be particularly well defined. The cohorts within this range are shown in Fig. \ref{fig-2}. We use linear regression to fit the following: $T_{ij}=\beta\log \tau_{ij}+\alpha$ to the data. The larger the $\beta$, the stronger is the dependence. In Fig. \ref{fig-4}(a) we provide the values of $\beta$ for the different ages and genders (filled symbols). Overall, the effect is strongest in the age range of $25-40$ year olds and for same-sex pairs. 

\begin{figure}
\includegraphics[scale=0.9]{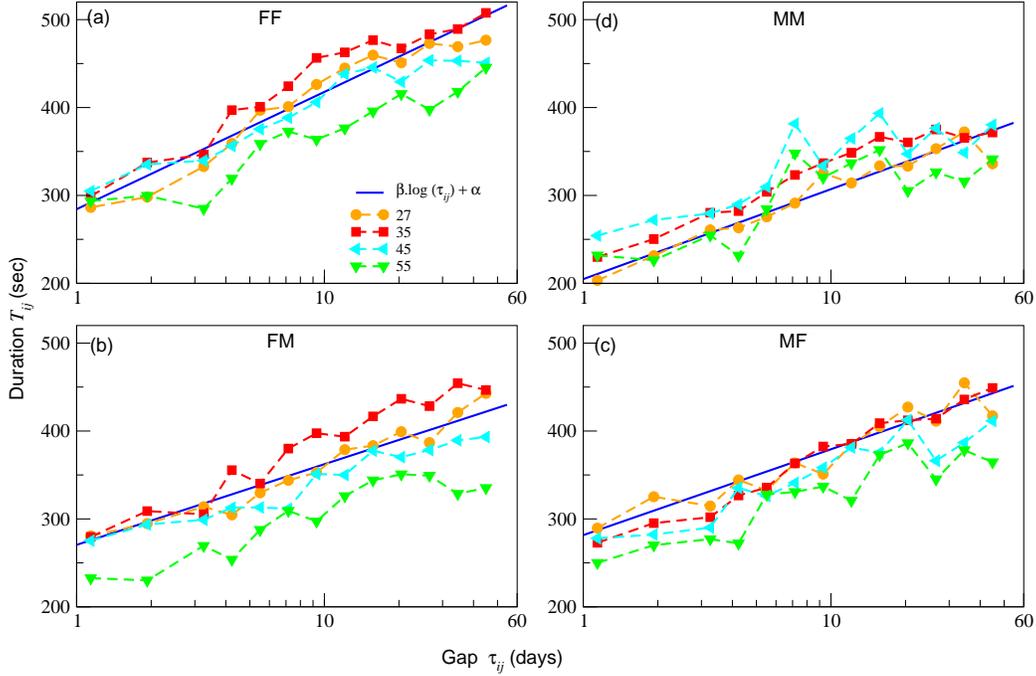}
\caption{Duration of succeeding call ($T_{ij}$) as a function of the gap ($\tau_{ij}$) for different ego-alter pairs belonging to the set $\mathcal{S}$. The curves shown with different symbols are obtained by log-binning the actual data corresponding to the pairs. The age cohorts of the ego are provided in the figure legend of (a). The plots correspond to the different sexes of the ego and the alter: female-female (FF) (a), female-male (FM) (b), male-female (MF) (c) and male-male (MM) (d). The solid line corresponds to a regression fit of the form: $\beta\log \tau + \alpha$, to the scatter corresponding the ego age cohort $35$.
}
\label{fig-2}
\end{figure}

\begin{figure}
\includegraphics[scale=1.0]{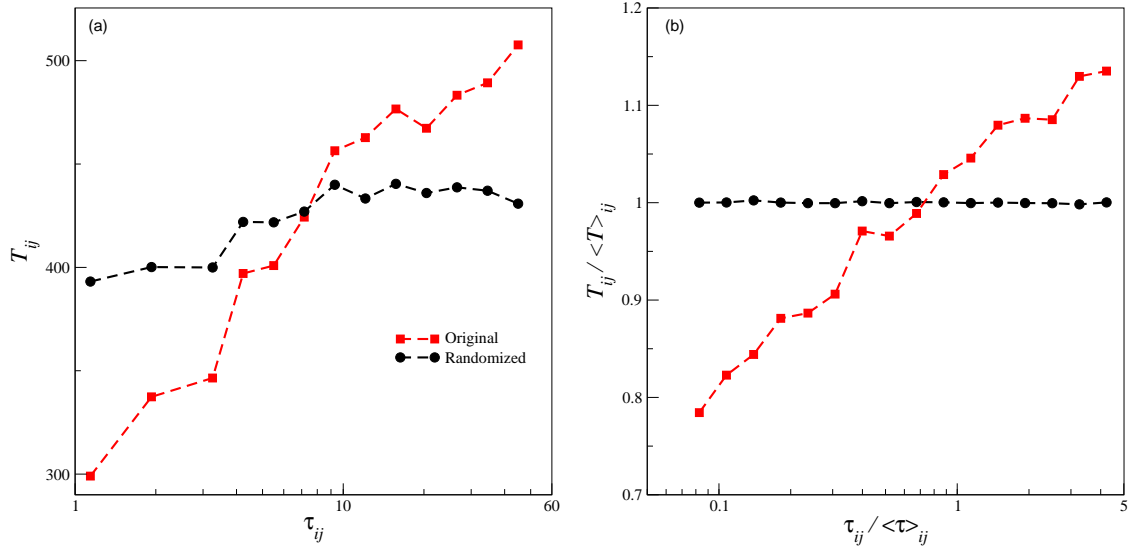}
\caption{(a) The red squares denote the duration of the succeeding call ($T_{ij}$) as a function of the gap ($\tau_{ij}$) for female-female pairs chosen from the set $\mathcal{S}$ with egos in the age cohort $35$. The black circles correspond to the data obtained from different realizations when the gaps for each pair are randomly shuffled. (b) The curves result from rescaling the data used in (a). For each pair $ij$, the gaps $\tau_{ij}$ are scaled by the average gap ${\langle\tau\rangle}_{ij}$ for the pair. Similar scaling is used for the durations.
}
\label{fig-3}
\end{figure}

\begin{figure}
\includegraphics[scale=1.0]{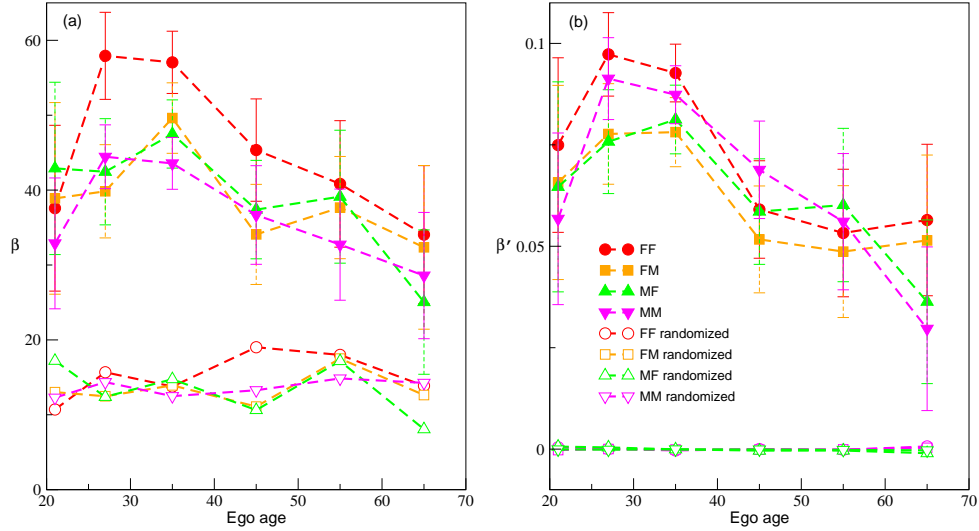}
\caption{(a) The slopes ($\beta$) of regression fits plotted as a function of the ego age for the original data and the randomized data.  The legends are same as in (b). (b) The regression slopes ($\beta'$) corresponding to the scaled data. The error bars span the $95\%$ confidence level.
}
\label{fig-4}
\end{figure}

The dependence of $T_{ij}$ on $\tau_{ij}$ as reflected in Fig. \ref{fig-2} results from accumulating the data from multiple sequences belonging to different ego-alter pairs. However, different pairs are expected to have their own idiosyncrasies, and as evident from Fig. \ref{fig-1}, the averages (gap and call duration) corresponding to different pairs follow unimodal distributions. Therefore, we first analyze the extent to which the properties of different pairs influence this dependence. For a set $\{(\tau_{ij},T_{ij})\}$ belonging to a given pair $ij$, we construct an ensemble of artificial sets $\{(\tau_{ij},T'_{ij})\}$, where, the $T'_{ij}$'s are obtained by randomly shuffling the original sequence of $T_{ij}$'s. In Fig. \ref{fig-3} (a) we illustrate the behaviour of the artificial data (black circles) for a particular case. The manufactured durations show a much weaker increase when compared to the original (red squares). We show the $\beta$'s resulting from the regression on the randomized data in Fig. \ref{fig-4} (a) (unfilled symbols). In general, the slopes for the randomized data are much lower, although different from zero. This comparison suggests that the correlations are truly present in the real data.

To extract the behaviour in a form that is independent of the characteristics of the ego-alter pairs, we scale the variables for a given pair with their corresponding averages. The dependence of the scaled variable $T_{ij}/\langle T\rangle_{ij}$ on $\tau_{ij}/\langle \tau\rangle_{ij}$ is shown in Fig. \ref{fig-3} (b) (red squares). The fact that the scaling extracts the correct nature of the correlations is evidenced when we scale the randomized data. The resulting curve (black circles) is flat and shows the absence of any correlation when the data are randomized. We employ a regression of the form: $T_{ij}/\langle T\rangle_{ij}=\beta'\log \left(\tau_{ij}/\langle \tau\rangle_{ij}\right)+\alpha'$ for the scaled data. In Fig. \ref{fig-4} (b) we plot the slopes $\beta'$. The figure clearly illustrates that the relationship between the scaled variables is qualitatively the same as that of the unscaled variables. Also, the scaled variables exhibit clear correlations, whereas the slopes for the randomized data are not different from zero. The relationship shows that for a given pair when the length of the gap is larger than the average gap, the duration of the successive call is larger than the duration of the average call. Conversely, if the gap is less than the average, then the duration also falls below the average.

To closely examine the nature of the ties we construct the distribution of alter age $-$ ego age for the pairs in $\mathcal{S}$. The distributions (SM) show that $\mathcal{S}$ is predominantly constituted by alters having the same sex as that of the ego and falling in the same age cohort. In general, the above pattern holds up to the age of $50$ year olds. For egos aged above $50$, peaks appears at an age separated by one generation. The preferred alters (those called most often) are mainly age peers. Because these dyads were in different geographical locations, they are unlikely to be spouses (indeed, most are same sex peers) and are more likely to be either friends, siblings or distant similar age kin (e.g. cousins). Above the age of 50 years, the double peak suggests that, in addition to peers, egos invest heavily in alters that are about a generation younger, most likely either children or nephews/nieces.

We further categorized the pairs in the set $\mathcal{S}$ based on the distance and the frequency of communication. First we divide the set into two groups, one consisting of pairs with distances ($d_{ij}$) smaller than $d_c$ and the other with distances larger than $d_c$. We consider the average gap as a proxy for the frequency of calling. Therefore, we again split each of these groups into two subsets based on whether the average gap ($\langle \tau\rangle_{ij}$) is less than $\tau_c$ or greater than $\tau_c$. We choose $d_c=50$ km which is larger than the spatial extension of the largest cities in the concerned country. Note that, in general, distance to alters is distributed according to an inverse power law \cite{onnela2011geographic}. We choose $\tau_c=12$, which is the most probable value of $\langle \tau\rangle_{ij}$ as can be seen in Fig. \ref{fig-1}. (See SM for the joint PDF of average gap and distance).

\begin{figure}
\includegraphics[scale=1.0]{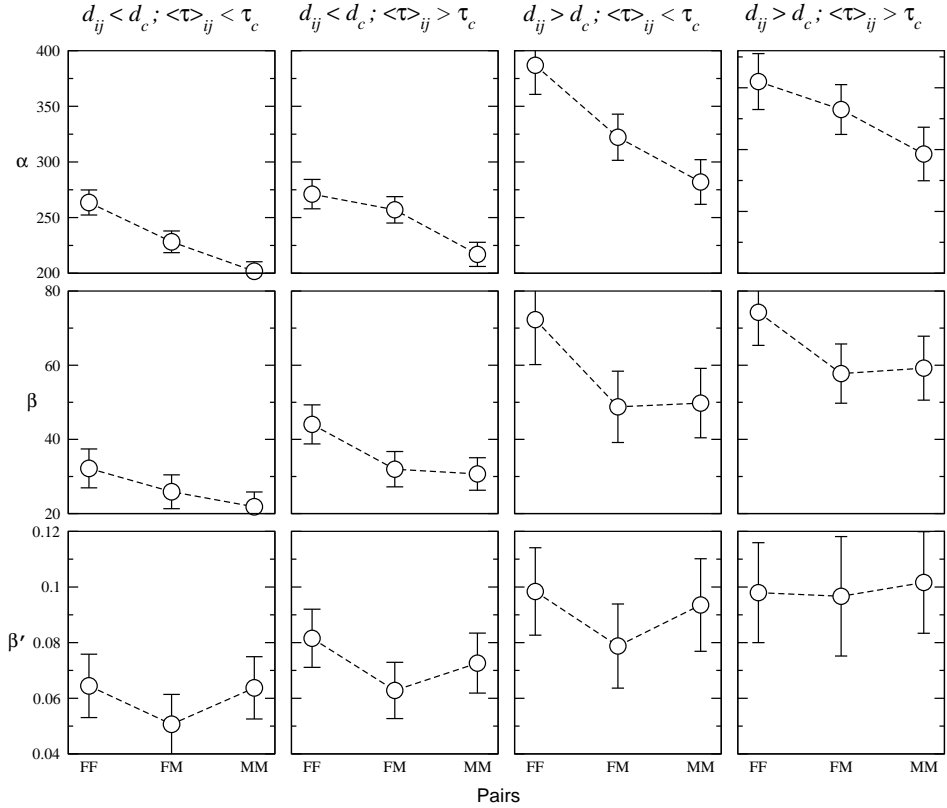} 
\caption{
The coefficients $\alpha$, $\beta$ and $\beta'$ resulting from regression fits when pairs in the set $\mathcal{S}$ are categorized into different subsets. A broad distinction into 
four groups (as indicated on the top of the columns) is done based on whether for a given pair the distance between the most common locations ($d_{ij}$) is lesser or greater than $d_c$; and whether average gap ($\langle\tau\rangle_{ij}$) is lesser or greater than $\tau_c$. A finer classification is made based on the gender of the individuals as indicated along the horizontal axis. Pairs are chosen irrespective of their age. The dashed line is a guide to the eye. (See SM for values without gender classification).  
}
\label{fig-5}
\end{figure}

In Fig. \ref{fig-5} we provide the values of the coefficients $\alpha$, $\beta$ and $\beta'$ for the fits to the data with the categorization as described in the previous paragraph. We obtain the coefficients with the data being further classified according to the gender of the individuals forming the pair. (See SM for the coefficients when pairs are analyzed irrespective of the gender). The coefficient $\alpha$ provides a basal value for the duration of the calls. The plot clearly indicates an increase in $\alpha$ with the increase in distance. However, variation with average gap is not significant. The fact that females are involved in longer calls than the males is also evident. For the $\beta$'s we observe a variation with distance very similar to $\alpha$. However, there is a marginal dependence on the average gap. These facts suggest that the reinforcement effect is stronger when the calling frequency is low and the distance of separation is large. The values of $\beta'$ reflects the fact that the observation regarding $\beta$'s  persist when the data is scaled. It also shows a strong gender homophily as the $\beta'$'s for same gender pairs appear to be larger compared to mixed gender pairs. This observation is consistent with the results shown in Fig. \ref{fig-4} (b).

\section*{Discussion}

Our main focus in the study has been with whether individuals adjust the time investment they make in relationships that matter to them when these are at risk. Our index of being `at risk' was a greater than the average gap between contacts (in this case phone calls). Friendships require constant time investment for their maintenance, and failure to match quite specific investment schedules leads inexorably to a rapid reduction in relationship quality \cite{roberts20106,sutcliffe2012relationships,roberts2015managing}. Our findings demonstrate a logarithmic increase in call duration with an increase in inter-call gap between a pair of individuals. In addition, we found that there is marked gender homophily in who calls who (and especially so in respect of same-age peers) and showed that the dependence of call duration on gap is stronger for (1) younger individuals (and especially so in $30$-year-olds) than for older people, (2) individuals separated by larger distances, (3) same-sex pairs, and (4) individual for whom the average inter-call gap is larger (marginal effect). Given that the frequency of contact determines the quality of a relationship \cite{roberts20106,sutcliffe2012relationships}, these results suggest that when individuals fail to contact each other frequently enough they compensate by devoting more time to the next call. This trade-off suggests that time acts as a form of social capital, and needs to be allocated to one's alters carefully so as to maintain relationships that are considered to be important. 

If we assume the complete or partial lack of face-to-face contact in such relationships (because dyads were in different geographical locations), then the lack of opportunity for face-to-face communication would naturally make it difficult to maintain relationship quality over time. The increase in the duration of the succeeding call after a long gap (time gap with the previous call) could thus serve as an act of relationship repair. The nature of the calls can also be inferred from the peaks (in multiples of $7$ days) in the distribution of gaps (Fig. \ref{fig-1} (a)). This suggests that the alters being called are among the set of emotionally close members of ego's network (somewhere within the inner core of five, normally contacted at least once a week, and the next layer out summing to $15$ that are called at least once a month) \cite{sutcliffe2012relationships}. 

These results parallel those noted in a number of animal species \cite{altmann2001baboon,dunbar1988maternal,moss1983relationships,wittig2005repair,moscovice2015welcome,aureli2002conflict,smith2011greetings}, suggesting that where relationships provide valuable benefits that influence fitness,  individuals may be anxious to ensure their stability and future persistence. The fact that animals respond in this way might be taken as {\it prima facie} evidence for the ability to foresee the consequences of behaviour or events. That humans can do this is, of course, no surprise. However, it does underscore the importance attached to social relationships, suggesting that when these are perceived as likely to become degraded through, for example, lack of opportunity to interact, we make special efforts to reinforce them. We imagine that this is true only for those core relationships within the inner layers of egocentric networks \cite{sutcliffe2012relationships} and that such an effect is unlikely to be true of the weaker relationships that populate the outer layers of egocentric networks where interaction rates are already very low \cite{sutcliffe2012relationships}. The innermost network layers, typically amounting to just $15$ people, are the ones that provide us with our principal sources of support and protection \cite{curry2013altruism}.

\section*{Materials and methods}
\label{Methods}

For this study we use an anonymized dataset containing details of 
mobile phone communication of subscribers of a particular operator in a European country over a period of $7$ consecutive months during the year $2007$. The details include full calling histories for the subscribers. Out of all the subscribers, the egos considered for our study are those whose age, gender and the most common location (location of the most accessed cell tower) are known and are active over the entire period of seven months. We use the the following criteria to select ego-alter pairs for whom the communication is sufficiently spaced over time. We consider the pairs for whom the maximum number of calls in any calendar month does not exceed $c_{max}=4$. Also, we take into account the number of different months ($m_{ij}$) over which a pair participates in calls. To focus on pairs where regularity of communication is observed we use $m_{ij}=7$.  Also, we consider only those pairs for which the distance between their most common locations, $d_{ij}$ is greater than zero. We also use a minimum threshold for the total talking time in the entire period, $t_{min}=30$ minutes. The role of $t_{min}$ is to ensure that for a given pair, the relationship would be significantly valued by both as indexed by the time invested.  We term the set of ego-alter pairs chosen by the above criteria ($c_{max}=4$, $m_{ij}=7$, $d_{ij}>0$ and $t_{min}=30$) as $\mathcal{S}$. In the SM we show that the qualitative nature of our findings is sufficiently robust and is not delicately dependent on the choice of the different thresholds to define $\mathcal{S}$. The total number of egos considered is around $400,000$. We group the egos (males and females appearing in almost equal ratios) into different age cohorts denoted by $21$ ($18$--$24$), $27$ ($25$--$30$), $35$ ($31$--$40$), $45$ ($41$--$50$), $55$ ($50$--$60$) and $65$ ($60$--$70$) years.

\newpage

\section*{SUPPLEMENTARY MATERIAL (SM)\\Absence makes the heart grow fonder: social compensation when failure to interact risks weakening a relationship}

\begin{figure}[h]
\includegraphics[width=9cm]{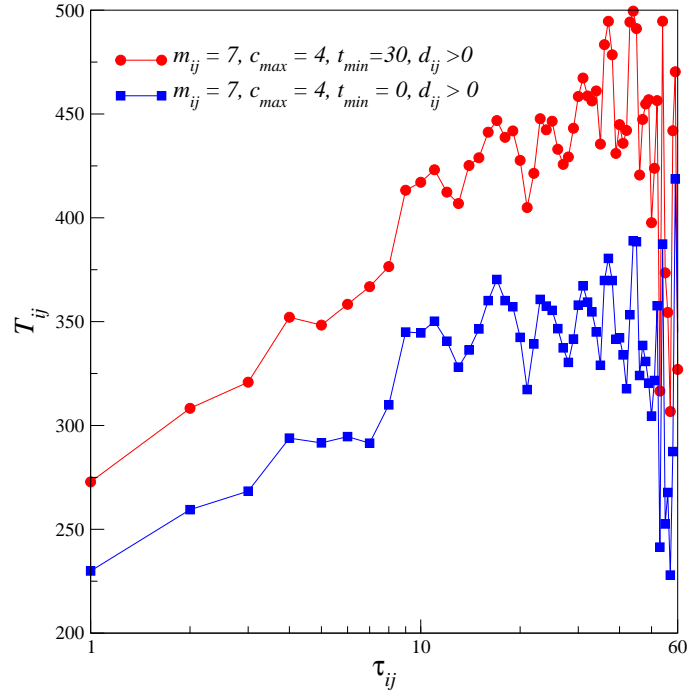}
\caption{{\bf Linear binnining of data for different thresholds ($t_{min}$).} Duration of succeeding call ($T_{ij}$) as a function of the gap ($\tau_{ij}$) for pairs belonging to the set $\mathcal{S}$ with the individuals aged between $25$ and $45$. Two different thresholds are used for the total aggregated duration in the $7$ month period -- $t_{min}=30$ minutes (red circles) and $t_{min}=0$ (no threshold) (blue squares). Linear binning is used for this plot unlike the log-binning used in the main text for reducing the fluctuation. The curves show a periodic drop in the call duration on multiples in $7$ days. This behaviour is related to the presence of peaks in the distribution of the gaps ($\tau_{ij}$). As such larger number of calls on the days at `weekends' result in a drop in the average duration of a call on those days. The equation we have used for regression primarily takes into account the dependence $T_{ij}\propto \log\tau_{ij}$. We have not taken into account the periodicity present in the data. 
}
\label{fig-1}
\end{figure}

\begin{figure}[h]
\includegraphics[width=14cm]{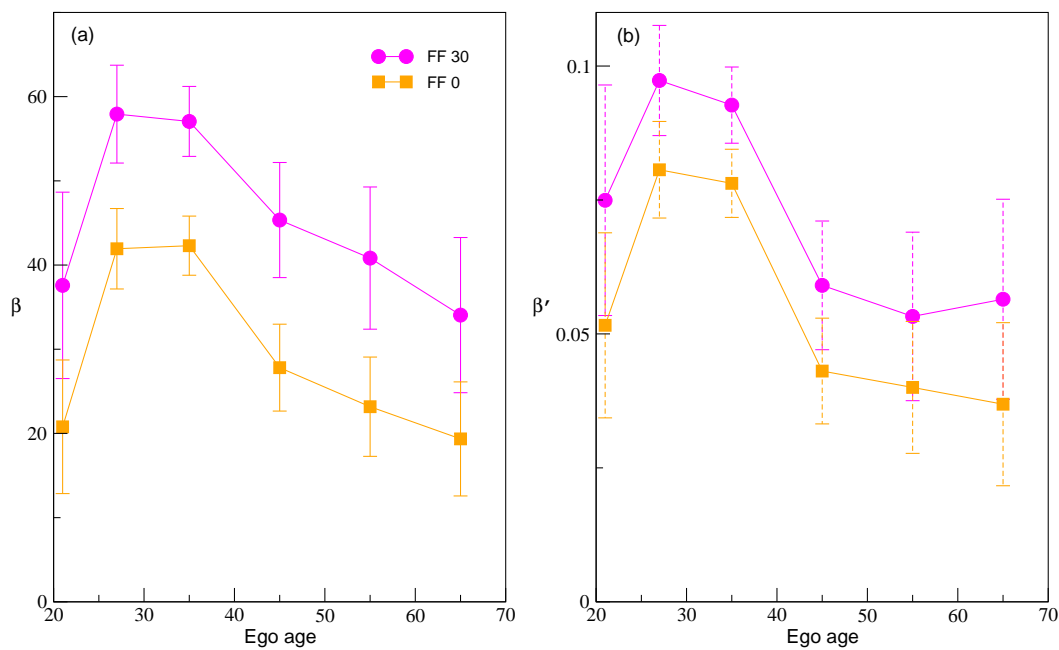}
\caption{{\bf Values of $\beta$ and $\beta'$ for different thresholds ($t_{min}$).} Comparison of the slopes of the regression fit (main text), $\beta$ (a) and $\beta'$ (b), for values of $t_{min}$ equal to $30$ minutes (circles) and zero (implying no threshold) (squares). For clarity only the values corresponding to female-female (FF) pairs chosen from set $\mathcal{S}$ are shown. The values of $\beta'$ show that even without a finite threshold on the calling duration, the values are different from zero.}
\label{fig-2}
\end{figure}

\begin{figure}[h]
\includegraphics[width=12cm]{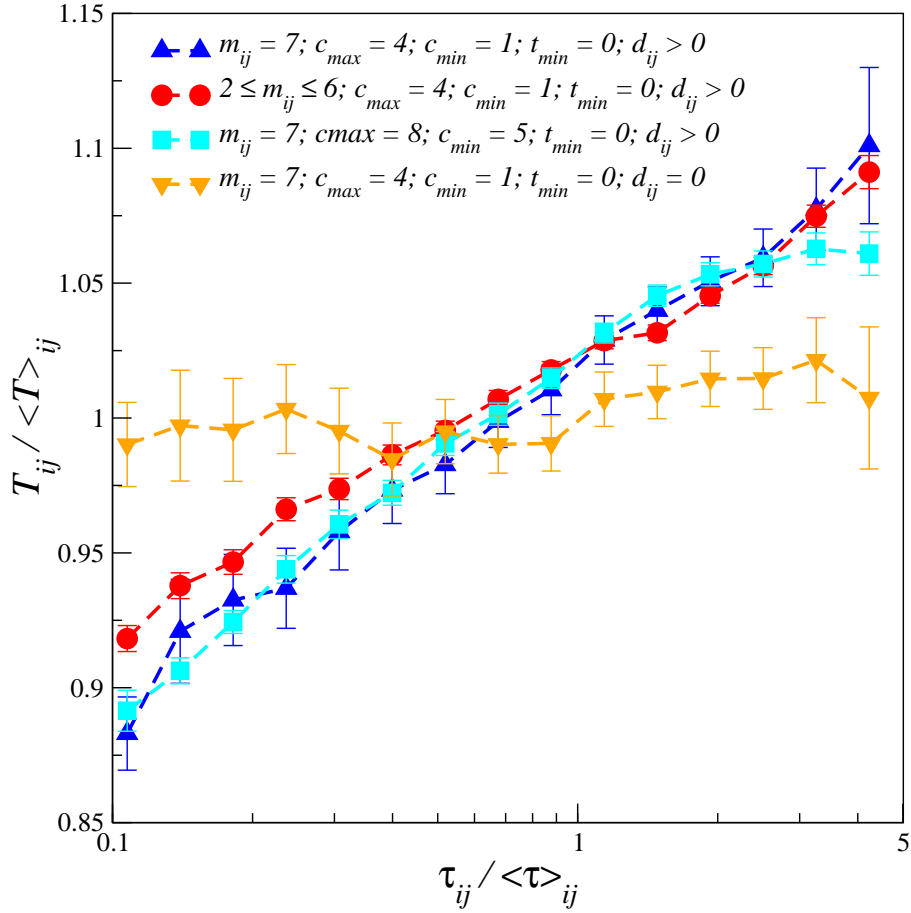}
\caption{{\bf Effect of using different filtering parameters.} The figure illustrates the dependence of the scaled durations, $T_{ij}/\langle T\rangle_{ij}$ on scaled gaps $\tau_{ij}/\langle\tau\rangle_{ij}$ when the pairs are chosen using one or more conditions ($m_{ij}$, $c_{max}$, $t_{min}$ and $d_{ij}$) which are different from those that are used to construct $S$. (We use an additional parameter $c_{min}$ which denotes the minimum number of calls between a pair in a any month. For $\mathcal{S}$, at least one call every month ensures $c_{min}=1$). The pairs are sampled irrespective of the age and gender of the individuals. The curve with blue up-triangles with no lower limit is set on the aggregated call duration in the $7$ month period ($t_{min}=0$, rest of the parameters being the same as in $\mathcal{S}$). The curve with red circles shows the behaviour when the condition $m_{ij}=7$ is relaxed. The pairs chosen are those that participate in calls in different months numbering from a minimum of $2$ to a maximum of $6$. The cyan squares show the case when the minimum number of calls in each of the $7$ months is set to $5$ and the maximum number is $8$, in contrast to the maximum being $4$ in $\mathcal{S}$. The orange down-triangles illustrate the case when pairs having the same most common locations are chosen. This curve is almost flat allowing for fluctuations.}
\label{fig-3}
\end{figure}

\begin{figure}
\includegraphics[width=15cm]{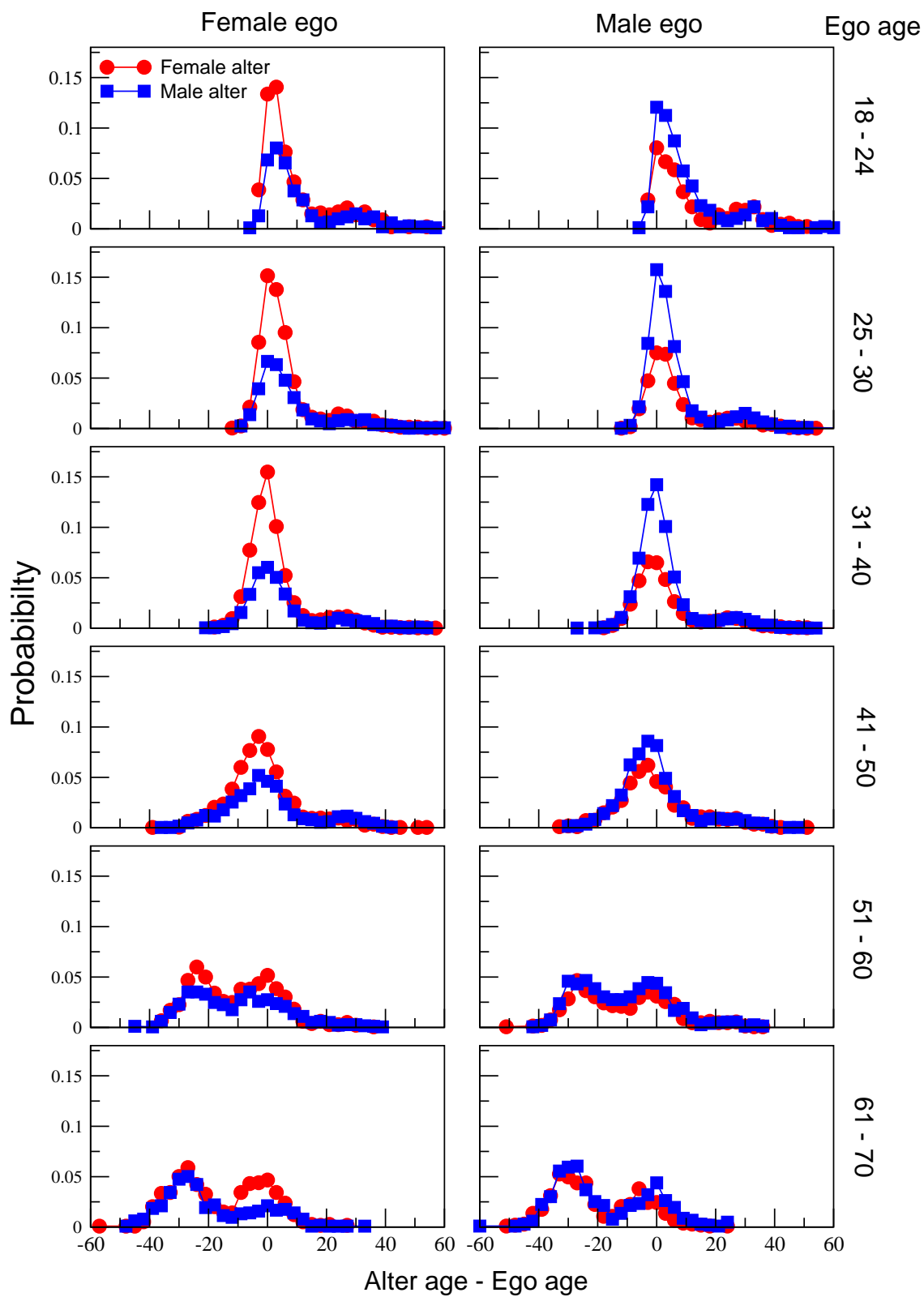}
\caption{{\bf Gender of individuals in the chosen pairs.} Distribution of (alter age$-$ego age) for female and male egos in different age cohorts with alters belong to the set $\mathcal{S}$.
}
\label{fig-4}
\end{figure}

\begin{figure}
\includegraphics[width=14cm]{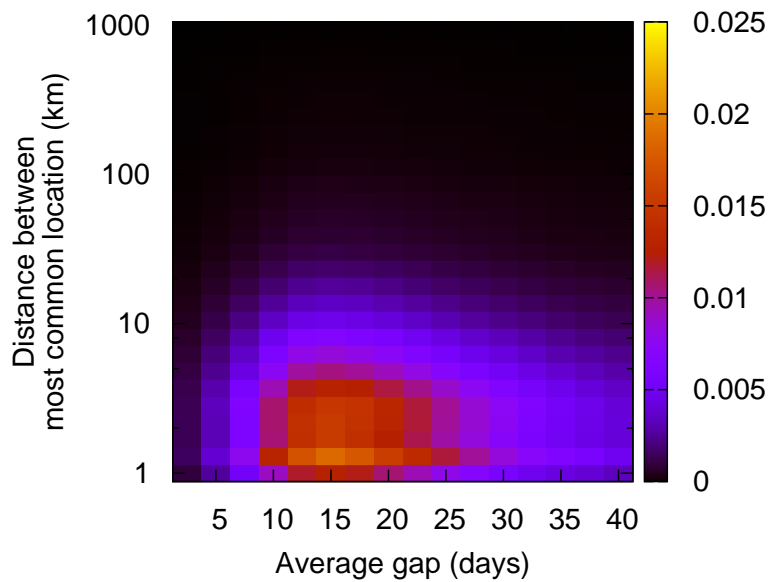}
\caption{{\bf Distribution of inter-call gap and geographical separation of individuals making up pairs.} Joint probabilty density function of average gap between pairs ($\langle\tau\rangle_{ij}$) and the distance between their most common location ($d_{ij}$). Pairs are chosen irrespective of age and gender.
}
\label{fig-5}
\end{figure}

\begin{figure}
\includegraphics[width=12cm]{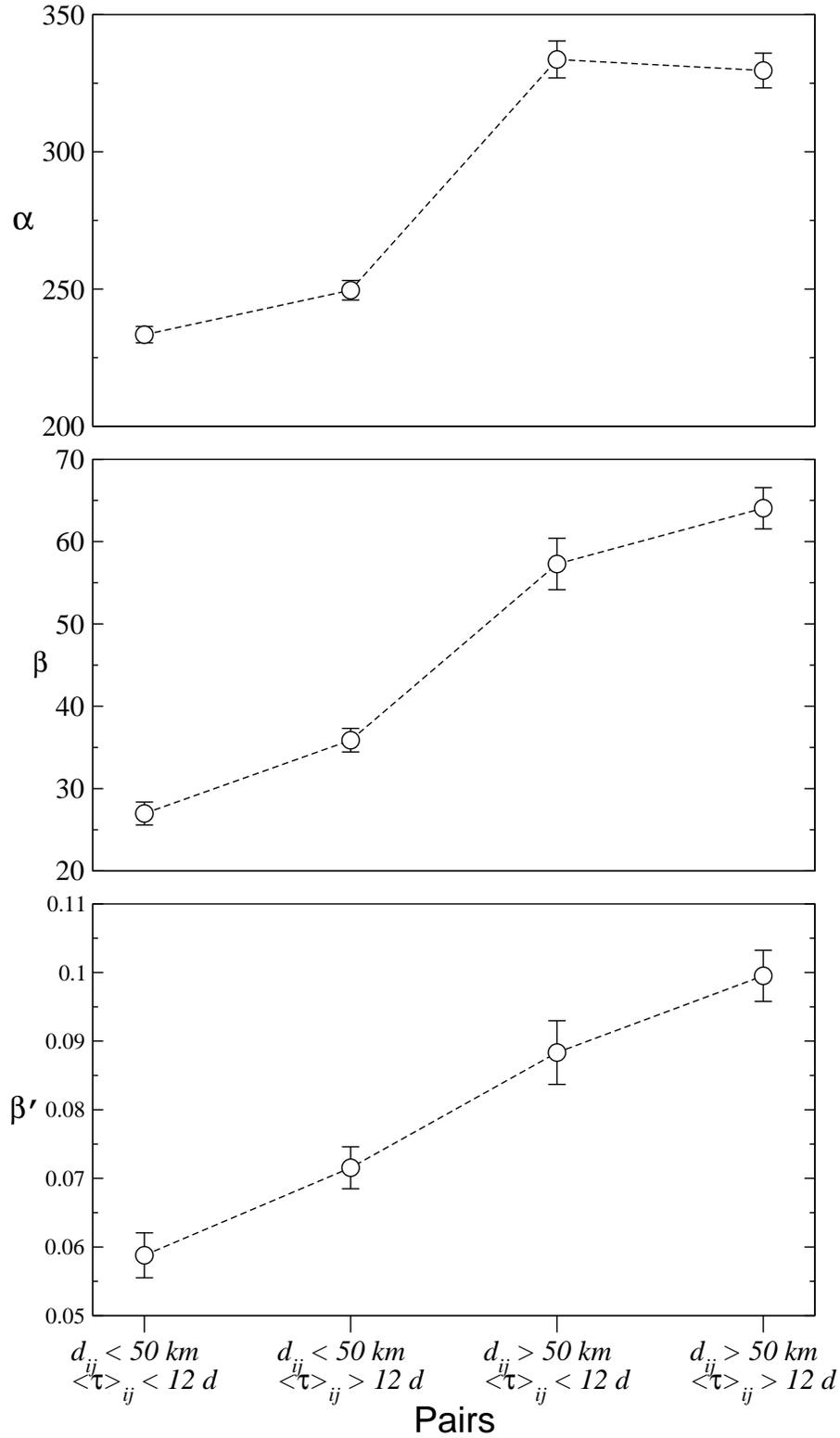}
\caption{{\bf Regression coefficients irrespective of the gender of individuals in the pairs.} Coefficients $\alpha$, $\beta$ and $\beta'$ corresponding four different categories based on distance between most common location ($d_{ij}$) and average gap  ($\langle\tau\rangle_{ij}$). Pairs are chosen irrespective of age and gender. The dashed line is a guide to the eye. (Also see Fig. 5 in the main text).
}
\label{fig-6}
\end{figure}

\end{document}